Юпиков О.А.
Севастопольский национальный технический университет

# ВЛИЯНИЕ ЧИСЛА ПРИЕМНЫХ КАНАЛОВ В АНТЕННОЙ СИСТЕМЕ С ФОКАЛЬНОЙ РЕШЕТКОЙ НА ЕЕ ЧУВСТВИТЕЛЬНОСТЬ

**Рассмотрена антенная система, состоящая из параболического зеркала и антенной решетки в качестве облучателя, описан метод моделирования данной системы и показано влияние количества приемных каналов на чувствительность всей системы. Расчеты проведены на примере системы APERTIF.**

**The antenna system with focal plane array is considered, the method for simulation of such system is described, and the influence of the number of receiving channels on the sensitivity of the system is shown.**

**Введение**

В настоящее время исследуют плотные решетки для использования в качестве облучателя зеркальной антенны [1—6]. Главным преимуществом плотных решеток над традиционными рупорными облучателями является то, что с их помощью можно достаточно точно сформировать требуемое амплитудно-фазовое распределение в апертуре зеркала для нескольких близко расположенных лучей, что не представляется возможным при применении рупорных облучателей для глубоких зеркал ($F/D<1$) [7-8]. Одновременное же (без поворота зеркала) формирование близко расположенных лучей открывает возможность увеличения поля обзора антенны без потери разрешающей способности, улучшения непрерывности поля обзора по сравнению с зеркальной антенной системой с рупором или группой рупоров в качестве облучателя.

Недостаток плотных антенных решеток состоит в сильном взаимном влиянии элементов решетки друг на друга, что может быть причиной значительного увеличения шумовой температуры системы [10-11]. На шумовые характеристики решетки также влияет согласование отдельных ее элементов с малошумящими усилителями. Но, в отличие от одиночной антенны, в решетке импеданс отдельных элементов зависит от формируемого луча, то есть от приложенных весовых коэффициентов. В статьях [9, 10] описана теория эффективного согласования элементов антенных решеток, которая использована при расчете шумовых характеристик рассмотренной антенной системы.

Очевидно, что при увеличении числа элементов решетки (а, следовательно, и числа приемных каналов системы), улучшится качество формирования лучей, но в то же время увеличится стоимость оборудования.

Цель данной работы — оценить зависимость основных характеристик приемной зеркальной антенны с фокальной решеткой от количества приемных каналов. Под приемным каналом понимается цепочка «элемент решетки – малошумящий усилитель (МШУ) – кабель – канал приемника».

Основной характеристикой радиотелескопа является его чувствительность, так как скорость обзора неба пропорциональна квадрату чувствительности [12]. Поэтому в качестве критерия качества системы выбран критерий чувствительности, и положено, что при уменьшении числа приемных каналов, чувствительность не должна уменьшиться более чем на 5% по сравнению с чувствительностью системы, в которой использованы все элементы решетки для формирования каждого луча.

Используя приведенный в статье анализ, можно компромисс между качеством системы и ее стоимостью.

Анализ проводился на примере разрабатываемой системы APRERTIF [13] для Вестерборкского радиотелескопа-интерферометра (WSRT), который находится в Нидерландах и состоит из 14 поворачивающихся 25-метровых рефлекторных антенн.

**1. Постановка задачи**

Каждая рефлекторная антенна интерферометра состоит из подвижного рефлектора, в фокусе которого установлена антенная решетка в закрытом корпусе. Сигнал, принимаемый каждым элементом решетки, усиливается и подается на приемник, который понижает частоту, фильтрует и оцифровывает сигнал, который в дальнейшем обрабатывается в цифровом виде.

При разработке APERTIF рассматривались несколько вариантов расположения функциональных блоков антенной системы. Первый из них, показанный на рис.1,а, является самым дешевым, т.к. в нем отсутствует транспортировка сигнала от каждого элемента решетки. Но этот вариант обладает рядом недостатков, наиболее существенные из которых – это, во-первых, создаваемые приемником и цифровыми схемами помехи, и, во-вторых, трудности охлаждения системы из-за ограниченного объема корпуса облучателя (выделяемая мощность показанных на рисунке блоков – несколько киловатт). Также существует ограничение на массу облучателя.

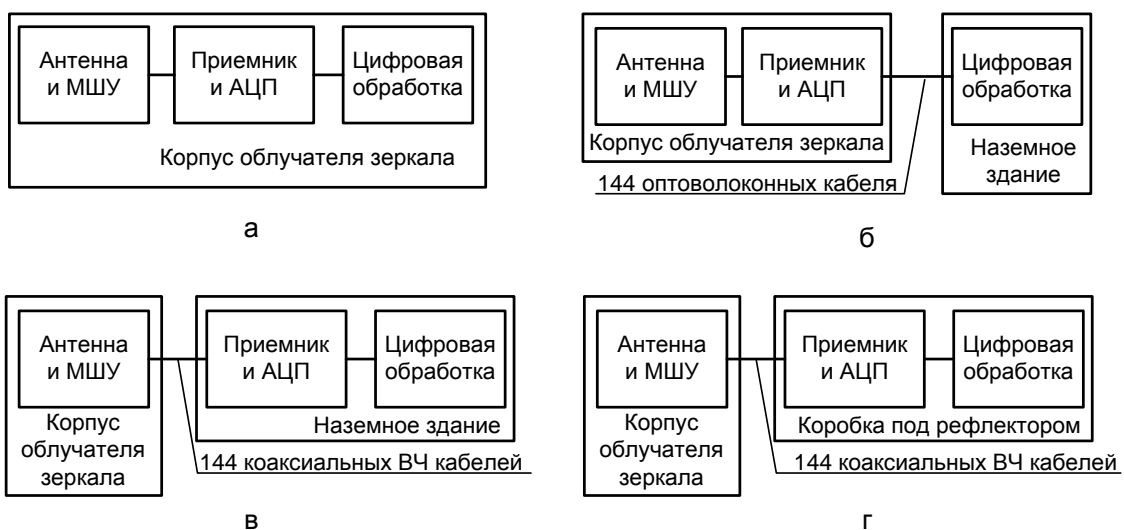

Рис. 1. Расположение функциональных блоков антенной системы

Во втором варианте (рис.1,б) в корпусе облучателя находится только антенная решетка с малошумящими усилителями (МШУ), подключенными к каждому ее элементу, и приемник, включающий в себя устройство оцифровки сигнала. Вся дальнейшая обработка сигнала происходит в наземном здании, расположенном возле антенны. В этом случае требования к электромагнитной совместимости внутри облучателя ниже, но к общей стоимости системы добавляется значительная стоимость оптоволоконных кабелей.

В третьем варианте (рис. 1, в) — в корпусе облучателя находится только антенная решетка с МШУ. Этот вариант свободен от главных недостатков первого варианта, а также уменьшена стоимость кабелей, т.к. в этом случае используется высокочастотный коаксиальный кабель вместо оптоволоконного. Но в то же время к кабелю предъявляются особые требования к фазовой стабильности и потерям на погонный метр. При данной конфигурации стоимость канала передачи сигнала от решетки к приемнику – примерно 100-150 евро на один элемент решетки (текущий прототип APERTIF имеет решетку со 144 элементами – по 72 элемента на каждую поляризацию, а расстояние от облучателя до здания у WSRT – около 60 м).

Четвертый вариант (рис. 1, г) аналогичен предыдущему, но приемник и цифровой процессор размещаются в коробке, находящейся под рефлектором. Это уменьшает длину кабелей и исключает их постоянный перегиб при вращении антенны, но вносит трудности монтажа и обслуживания.

Таким образом, наилучшим решением по отношению (надежность+простота)/стоимость был выбран третий вариант. Для достижения описанной во введении цели поставим задачу следующим образом: определить, сколько элементов решетки можно исключить *из процесса формирования луча*, чтобы чувствительность всей приемной антенной системы уменьшилась не более чем на 5%. Заметим, что уменьшать общее количество элементов в решетке нельзя, так как 1) для качественного формирования требуемого амплитудно-фазового распределения в апертуре зеркала требуется минимальное расстояние между элементами решетки, и 2) максимальный угол сканирования зависит от геометрических размеров решетки.

**2. Общий алгоритм решения задачи**

Решение поставленной задачи было разделено на следующие этапы:
1) электромагнитное моделирование антенной решетки, то есть расчет распределения токов в *металлической структуре* решетки, входного сопротивления и комплексных характеристик направленности (ХН) отдельных ее элементов;

2) расчет ХН элементов решетки после отражения излученного электромагнитного поля от зеркала (вторичных ХН);

3) микроволновое моделирование системы (см. подраздел 2.3) и расчет оптимальных по критерию максимальной чувствительности весовых коэффициентов для формирования лучей (рис.4);

4) обнуление тех из весовых коэффициентов, модуль которых меньше заданного значения $\Delta W_{trshld}$, где $\Delta W_{trshld}$ — порог обнуления, который изменяется от минус 30 до 0 дБ (см. пункт 6);

5) расчет результирующей ХН решетки, вторичной ХН, шумовых характеристик системы, которые используются в дальнейшем для расчета чувствительности при заданном наборе весовых коэффициентов;

6) повтор пунктов 4-5 для диапазона значений $\Delta W_{trshld}$ от минус 30 до 0 дБ.

Опишем каждый пункт приведенного выше алгоритма более подробно.

## 2.1. Расчет металлической структуры антенной решетки

Металлическая структура антенной решетки рассчитывалась методом характеристических базисных функций (CBFM) [14-16]. Этот метод представляет собой расширение метода моментов, и разработан специально для конечных псевдопериодических структур [17]. Он позволяет довольно точно рассчитывать распределение токов в металле структуры, но в то же время имеет значительно меньшие требования к объему памяти и вычислительной мощности компьютера, чем чистый метод моментов. Это позволяет проводить электродинамический расчет больших структур, которые на сегодняшний день не представляется возможным рассчитать при помощи чистого метода моментов, но, в то же время достаточно малы, чтобы использовать подход бесконечных решеток [17].

В результате расчета металлической структуры, используя описанный в [17] метод, имеем матрицу сопротивлений решетки **Z** (сопротивление между портами всех 144 элементов решетки), а также комплексные ХН каждого ее элемента с учетом окружения этого элемента.

## 2.2. Расчет вторичных диаграмм направленности

Вторичные ХН, то есть ХН, сформированные после отражения поля от зеркала, рассчитывались через первичные ХН элементов решетки при помощи программы GRASP9, в которой используется метод физической оптики (PO) и геометрическая теория дифракции (GTD) [18].

Вторичные диаграммы направленности (ДН) по мощности для углового и центрального элементов решетки, рассчитанные для частоты 1,42 ГГц, показаны на рис. 2. Здесь приведены нормированные ДН для случая, когда зеркало облучается одним элементом решетки (в присутствии остальных элементов).

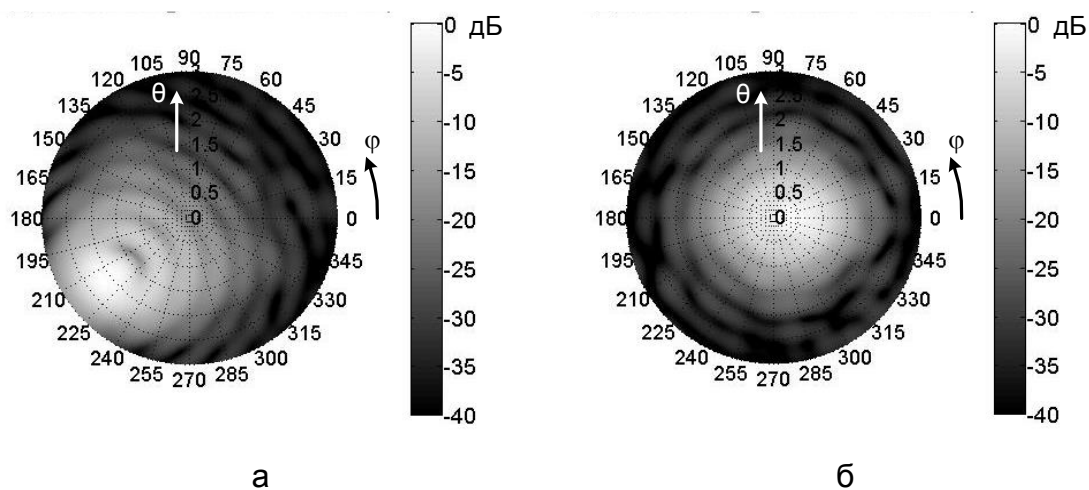

а                                      б

Рис. 2. Вторичные ДН для элементов решетки: углового элемента (а) и элемента, близкого к центру решетки (б)

## 2.3. Микроволновое моделирование антенной системы и расчет оптимальных по критерию максимальной чувствительности весовых коэффициентов

Далее металлическая структура решетки представлялась как $N$-полюсник, генерирующий сигнальные и шумовые волны. К его портам подключалось микрополосковое устройство питания [19] и малошумящие усилители, и проводилось микроволновое моделирование полученной структуры [20].

Структура традиционной антенной системы с фокальной решеткой приведена на рис. 3. Она условно состоит из двух блоков: 1) зеркало, элементы решетки с устройством их питания и малошумящие усилители (МШУ); 2) формирователь луча (управляемые аттенюаторы с фазовращателями и сумматор, как правило, цифровые). Разделение на такие блоки целесообразно с целью нахождения оптимальных весовых коэффициентов $w_n$ по критерию максимальной чувствительности.

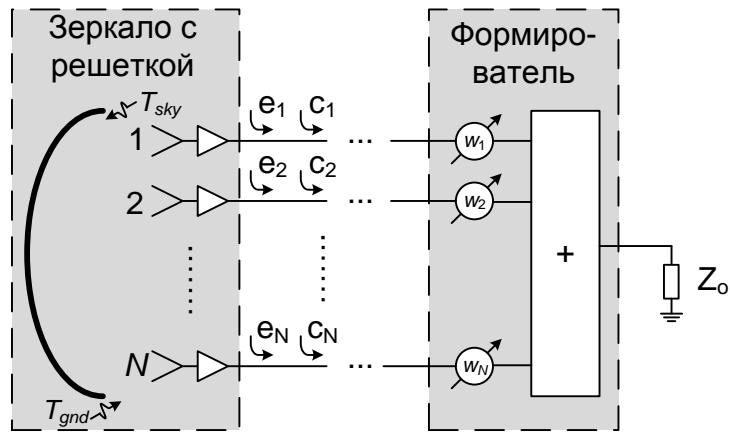

Рис. 3. Структура антенной системы и матрицы, используемые для расчета весовых коэффициентов.

Блок зеркала с решеткой и МШУ характеризуется (см. рис. 3) шумовой корреляционной матрицей между выходами элементов решетки

$$\mathbf{C} = \begin{pmatrix} c_{11} & c_{12} & \ldots & c_{1N} \\ c_{21} & c_{22} & \ldots & \vdots \\ \vdots & \vdots & \ddots & \vdots \\ c_{N1} & \ldots & \ldots & c_{NN} \end{pmatrix}$$

и вектором **e**, элементы которого содержат сигнальные волны [17, стр. 163-166] с выхода каждого элемента решетки (в отсутствии шумов) при приеме падающей волны с какого-либо определенного направления:

$$\mathbf{e} = \begin{pmatrix} e_1 & e_2 & \ldots & e_N \end{pmatrix}^T,$$

Для расчета весовых коэффициентов может применяться, например, метод согласования по полю. Но он не позволяет получить лучшее отношение сигнал/шум на выходе системы, т. к. шумы на выходе каждого из каналов коррелированны (высокий коэффициент корреляции) отчасти из-за приема внешних шумов, отчасти из-за сильной взаимной связи между элементами плотной решетки. Коэффициенты возбуждения решетки, оптимальные по критерию максимальной чувствительности, можно рассчитать, используя следующую матричную формулу [21]:

$$\mathbf{w} = \mathbf{C}^{-1}\mathbf{e},$$

в которой вектор **e** содержит сигналы с выхода каждого элемента решетки при приеме падающей волны с направления, для которого оптимизируются **w**.

Что касается максимального угла сканирования, то он определяется требованиями к размеру поля обзора телескопа. Для системы APERTIF – это 8 квадратных градусов, внутри которого находится 37 лучей. Максимальный угол сканирования при таком поле обзора равен приблизительно 1,5°.

Так как решетка является фокальной, то при приеме сигнала с некоторого направления основная часть энергии сосредотачивается (фокусируется) только лишь на небольшом участке решетки. Ограниченные размеры решетки также накладывают ограничения на максимальный угол сканирования.

На рис. 4 показаны модули найденных оптимальных по критерию максимальной чувствительности коэффициентов возбуждения элементов Вивальди фокальной решетки размером 8x9 на частоте 1420 МГц для осевого направления $\theta=0°$ (рис. 4 а, б) и крайнего направления $\theta=1,5°$ (рис. 4 в, г). На рисунке квадратиками обозначены элементы решетки одной поляризации, расположенные в соответствующих реальной решетке позициях.

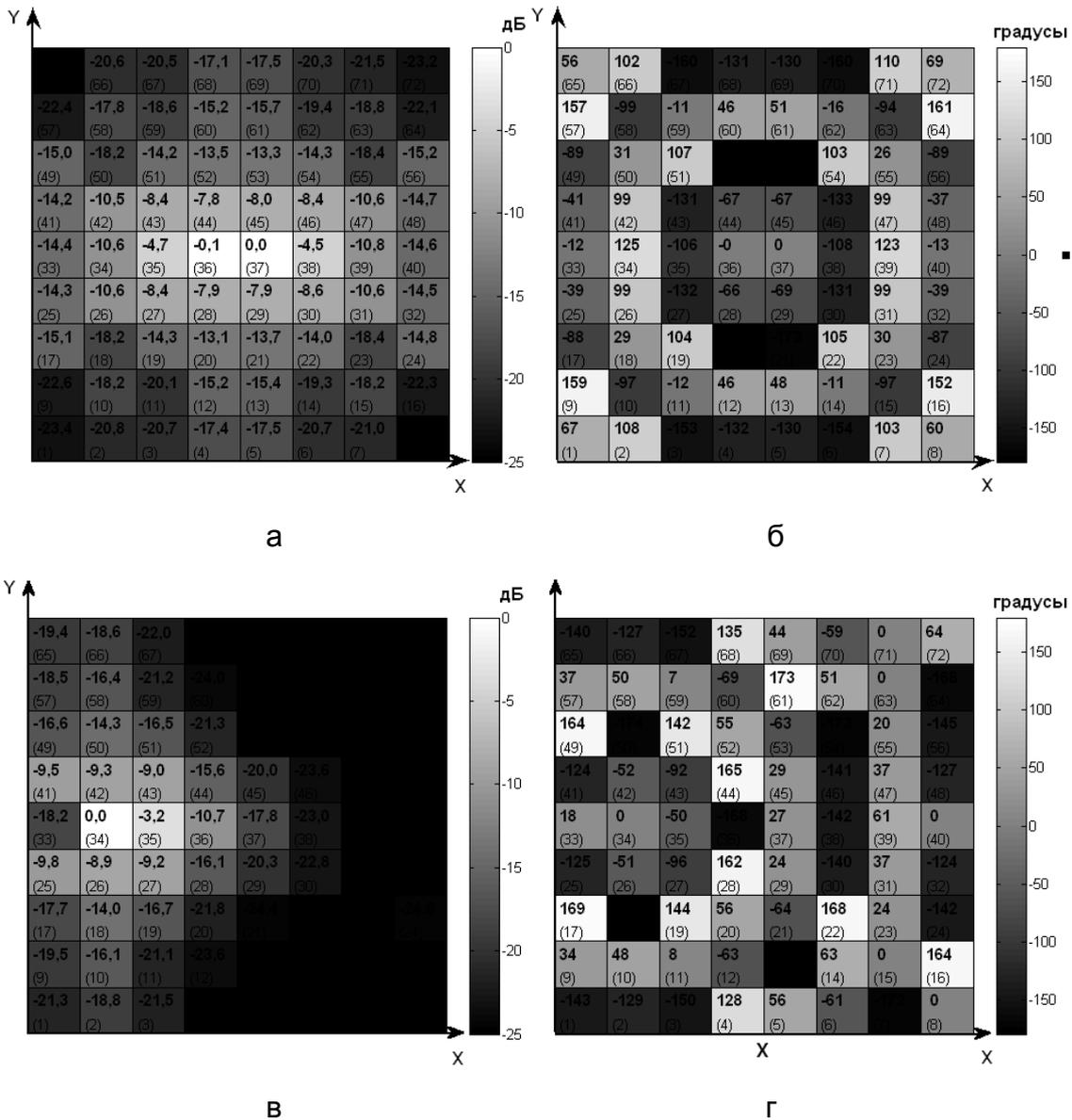

Рис.4. Весовые коэффициенты элементов фокальной решетки для центрального (а, б) и сканирующего на 1,5 градуса (в, г) лучей: их модули, выраженные в дБ (а, в), и аргументы, выраженные в градусах (б, г)

Из рисунка следует, что сильно возбуждены ($|w| > -10$ дБ) не более половины элементов. Следовательно, для выбранного направления сканирования не обязательно использовать все элементы решетки, а значит, можно сократить число каналов приема, тем самым уменьшить стоимость каналов передачи сигналов от элементов решетки к приемнику и стоимость самого приемника. При этом переключение оставшихся каналов приема между подрешетками, на которые падает основная часть энергии, может осуществляться при помощи высокочастотных мультиплексоров.

При обнаружении слабых сигналов на уровне шумов одним из наиболее важных параметров антенной системы является чувствительность $A_{эфф}/T_{сис}$, где $A_{эфф}$ – эффективная площадь антенны, $T_{сис}$ – общая шумовая температура всей системы. Мультиплексоры, включенные в систему, будут добавлять в нее шум (увеличивать $T_{сис}$), поэтому их следует включать после МШУ, то есть между МШУ и приемником. Таким образом, сокращению могут подлежать только каналы приемника, а также кабели от облучателя к формирователю, но не МШУ.

## 2.4. Определение чувствительности системы для разных величин порога $\Delta W_{trshld}$

После расчета ХН отдельных элементов решетки и весовых коэффициентов обнулялись те из весовых коэффициентов, модуль которых меньше заданного порога $\Delta W_{trshld}$, дБ (пункт 4 приведенного выше алгоритма), и для полученного набора весовых коэффициентов рассчитывались результирующие ХН решетки, результирующие вторичные ХН, шумовые характеристики системы и ее чувствительность (пункт 5 алгоритма). При этом предполагалось, что все элементы решетки, и активные (возбужденные), и неактивные (с нулевыми весовыми коэффициентами), нагружены на МШУ. Данные действия повторялись для диапазона значений $\Delta W_{trshld}$ (пункт 6 алгоритма).

Результаты расчета чувствительности антенной системы для двух лучей, и соответствующие им коэффициент использования поверхности (КИП), шумовая температура системы и коэффициент потерь, учитывающий эффект «затекания» энергии за края зеркала, показаны на рис. 5. На графиках по оси абсцисс отложен порог обнуления весовых коэффициентов $\Delta W_{trshld}$, по левой оси ординат — одна из характеристик антенной системы (линия с круглыми маркерами) и по правой оси ординат — количество возбужденных элементов (линия с треугольными маркерами). Цифрами 1 и 2 обозначены характеристики системы для осевого и сканирующего на 1,5° луча соответственно. Цифрами 3 и 4 обозначены кривые зависимости количества активных элементов от порога $\Delta W_{trshld}$ для осевого и сканирующего на 1,5° луча соответственно.

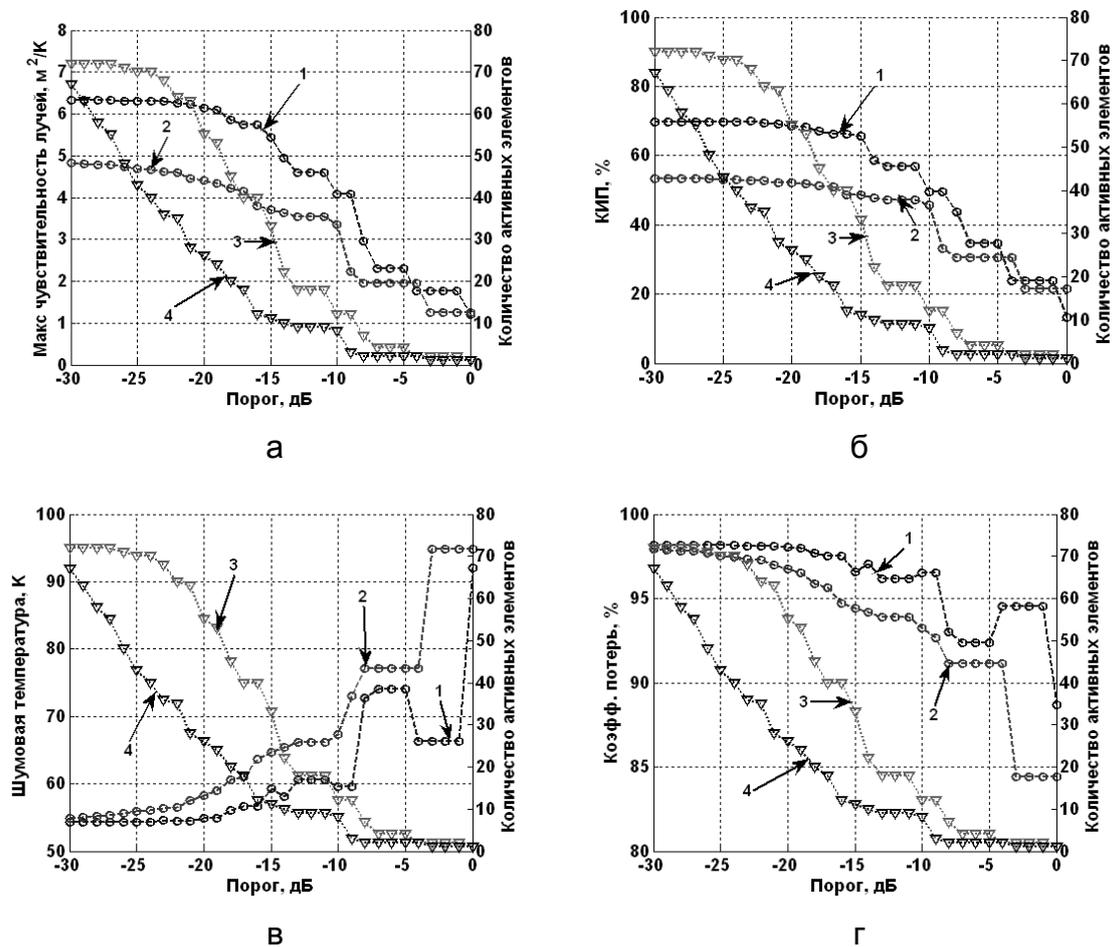

Рис.5. Зависимость основных характеристик антенной системы и количества активных элементов решетки от порога обнуления весовых коэффициентов $\Delta W_{trshld}$ для двух лучей: а – чувствительность лучей, б – КИП, в – шумовая температура системы, г – коэффициент потерь из-за эффекта «заливания» энергии за края зеркала.

**Выводы**

По приведенным графикам видно следующее. Во-первых, для рассматриваемого типа системы требуется большее количество элементов, так как для крайнего луча их не достаточно, что видно по рис. 4 в (бо́льшая часть принятой и отраженной зеркалом энергии источника не перехватывается решеткой), а значит, теряется чувствительность. Во-вторых, для данной системы можно ограничиться примерно 45 каналами приема для осевого направления, вместо 72 (количество элементов в решетке для одной поляризации), и при этом чувствительность системы уменьшается не более чем на 5 %. Для крайнего луча (угол сканирования 1,5°) требуемое число каналов составляет 35.

Таким образом, показано, что число приемных каналов может быть уменьшено почти на 40 % при незначительном (5 %) уменьшении чувствительности всей антенной системы.

**Список литературы**

1. Fisher J., Bradley R.F. Full Sampling Array Feeds for Radio Telescopes // Proc. of the SPIE, Radio Telescopes .— 2000 .— vol. 4015 .— P. 308 — 318.

2. Ivashina M.V., bij de Vaate J.G., Braun R., Bregman J.D. Focal Plane Arrays for large Reflector Antennas: First Results of a Demonstrator Project // Proc. of the SPIE, Astronomical Telescopes and Instrumentation .— Glasgow, UK, June 2004.

3. Cavallo D., Neto A., Gerini G., Toso G. On the Potentials of Connected Slots and Dipoles in the Presence of a Backing Reflector // 30$^{th}$ ESA Antenna Workshop on Antennas for Earth Observation, Science, Telecommunication and Navigation Space Missions .— 2008.

4. Del-Rно C., Betancourt D. Multi-Beam Applications of CORPS-BFN: Reflector Antenna Feeding System // 30$^{th}$ ESA Antenna Workshop on Antennas for Earth Observation, Science, Telecommunication and Navigation Space Missions .— 2008.

5. DeBoer D., Gough R., Bunton J., Cornwell T. et. al. The Australian SKA Pathfinder: A High-Dynamic Range Wide-Field of View Survey Telescope Array // Proceedings of the IEEE. — 2009. — vol. 97, issue 8. — P. 1507—1521.

6. Arts M., Ivashina M., Iupikov O., Bakker L., van den Brink R. Design of a Low-Loss Low-Noise Tapered Slot Phased Array Feed for Reflector Antennas // Proc. of EuCAP2010. — March 2010. — P. 1—5.

7. M. V. Ivashina, C.G.M. van 't Klooster, "Focal Fields in Reflector Antennas and Associated Array Feed Synthesis for High Efficiency Multi-Beam Performances", 25th ESA Antenna Workshop on Satellite Antenna Technology, Noordwijk, The Netherlands, September, 2002.

8. Ivashina M.V., Mou Kehn M.Ng., Kildal P.-S. Optimal number of elements and element spacing of wide-band focal plane arrays for a new generation radio telescope // Proc. EuCAP2007. — Edinburgh, UK, Nov. 2007. — P. 1 — 7.

9. Warnick K. F., Jensen M. A. Optimal noise matching for mutually-coupled arrays // IEEE Transactions on Antennas and Propagation. — June 2007. — Vol. 55. — P. 1726 — 1731.

10. Ivashina M.V., Maaskant R., Woestenburg E.E.M. Equivalent System Representation to Model the Beam Sensitivity of Receiving Antenna Arrays // IEEE Antennas Wireless Propag. Letter (AWPL). — Oct. 2008. — Vol. 7. — P. 733 — 737.

11. Maaskant, R. Bekers, D.J., Arts, M.J., van Cappellen, W.A., Ivashina, M.V., 'Evaluation of the Radiation Efficiency and the Noise Temperature of Low-Loss Antennas,' *Antennas and Wireless Propagation Letters, IEEE*, Dec. 2009, vol. 8, p. 1166-1170. Number of citations: 14. (without self-citations: **12**)

12. Hall P.J. An SKA Engineering Overview // SKA memos: memo 91. — Sep. 2009. — http://www.skatelescope.org/PDF/memos/memo_91.pdf. — 03.03.2011.

13. Aprtif. — http://www.astron.nl/general/apertif/apertif. — 04.01.2011.

14. Prakash V., Mittra R. Characteristic basis function method: A new technique for efficient solution of method of moments matrix equations // Micr. Opt. Technol. — Jan. 2003. — Vol. 36. — P. 95 — 100.

15. Maaskant R., Mittra R., Tijhuis A. G. Application of trapezoidalshaped Characteristic Basis Functions to arrays of electrically interconnected antenna elements // International Conference on Electromagnetics in Advanced Applications (ICEAA), Torino. — Sep. 2007.

16. Maaskant R., Mittra R., Tijhuis A. Fast Analysis of Large Antenna Arrays Using the Characteristic Basis Function Method and the Adaptive Cross Approximation Algorithm // IEEE Trans. on Ant. and Propagat. — Nov. 2008. — Vol.56, Issue 11, № 1. — P. 3440 — 3451.

17. Maaskant R. Analysis of Large Antenna Systems: PhD Thesis: 07.06.2010 / Rob Maaskant. — Rotterdam, 2010. — 273 p.

18. Pontoppidan K. GRASP9: Technical Description // TICRA Engineering Consultants, Tech. Rep. — 2005. — 403 p. (http://www.ticra.com/).

19. Ivashina M.V. Redkina E.A., Maaskant R. An Accurate Model of a Wide-Band Microstrip Feed for Slot Antenna Arrays // The 2007 IEEE International Symposium on Antennas and Propagation – 2007. — June 10-15, 2007. — P. 1953 — 1956.

20. Maaskant R., Yang B. A combined electromagnetic and microwave antenna system simulator for radio astronomy // EuCAP 2006 (DOI 10.1109/EUCAP.2006.4839073). — 2006. — P. 1 — 4.

21. Lo Y.T., Lee S.W. Optimization of directivity and signal-to-noise ratio of an arbitrary antenna array // IEEE Trans. — Aug., 1966. — Vol.54, № 8. — P. 1033 — 1045.

**Ключевые слова:** фокальная антенная решетка, чувствительность антенной системы, число приемных каналов.

**Key words:** focal plane array, antenna system sensitivity, number of the receiving channels.

Acknowledgement

This work has been supported by the Netherlands Institute for Radio Astronomy ASTRON, and conducted during Iupikov's visit to ASTRON during 2008-2009 under the supervision of Drs. Ivashina, Maaskant and Cappellen.